\documentstyle[twoside,amssymb,12pt]{article}

\newcommand{\nc}{\newcommand}
\nc{\la}{\lambda} \nc{\alf}{\alpha}  \nc{\T}{\Theta}
\nc{\tht}{\theta}  \nc{\be}{\beta}  \nc{\eps}{\epsilon} \nc{\z}{\zeta}
\nc{\ga}{\gamma}  \nc{\De}{\Delta}  \nc{\Ga}{\Gamma}  \nc{\vphi}{\varphi}
\nc{\de}{\delta} \nc{\si}{\sigma}  \nc{\ka}{\kappa}   \nc{\Si}{\Sigma}
\nc{\om}{\omega}  \nc{\qq}{\quad\quad}                \nc{\Om}{\Omega}
\nc{\nf}{\infty}   \nc{\dl}{\mathop{\smash{\cal L}}}  \nc{\black}{\rule{3mm}{3mm}}
\nc{\ra}{\rightarrow}  \nc{\ol}{\overline}  \nc{\und}{\underline}
\nc{\beq}{\begin{equation}}  \nc{\pt}{\partial}
\nc{\eeq}{\end{equation}}
\nc{\beqa}{\begin{eqnarray}}  \nc{\dst}{\displaystyle}
\nc{\eeqa}{\end{eqnarray}} \nc{\nnb}{\nonumber}
\nc{\bs}{\backslash}        \nc{\mb}{\mathbb}  \nc{\wdt}{\widetilde}

\newcounter{muni}
\newenvironment{remunerate}{\begin{list}{{\rm \arabic{muni}.}}
{\usecounter{muni}
\setlength{\leftmargin}{0pt}\setlength{\itemindent}{38pt}}}{\end{list}}

\nc{\brm}{\begin{remunerate}}   \nc{\erm}{\end{remunerate}}

\newtheorem{nth}{Proposition}   \newtheorem{nlem}{Lemma}
\nc{\stg}{\mathop{\smash{*}}}
\nc{\st}{\mathop{\smash{\delta}}}
\nc{\barr}{\begin{array}}   \nc{\earr}{\end{array}}   \nc{\dg}{\dagger}
\nc{\mtvb}{\mathversion{bold}}   \nc{\mtvn}{\mathversion{normal}}

\date{}
\begin{document}\begin{titlepage}
\begin{flushright} 
LPTHE 02-55 \\ 
hep-th/0212011 \\ 
November 2002 
\end{flushright} 
\vskip 1.0truecm
\centerline{\large \bf HIGHER CONSERVATION LAW FOR}
\centerline{\large \bf THE MULTI-CENTRE METRICS}
\vskip 1.0truecm 
\centerline{\bf Galliano Valent} 
\vskip 1.0truecm 

\centerline{\it LPTHE}
\centerline{\it Laboratoire de Physique Th\'eorique et des 
Hautes Energies,} 
\centerline{\it Unit\'e associ\'ee au CNRS UMR 7589}
\centerline{\it 2 Place Jussieu, 75251 Paris Cedex 05, France} 
\vskip 1.0truecm  \nopagebreak 

\begin{abstract}
The multi-centre metrics are a family of euclidean solutions of the empty space 
Einstein equations with self-dual curvature. For this full class, we determine 
which metrics do exhibit an extra conserved quantity quadratic in the momenta, 
induced by a Killing-St\" ackel tensor. Our results bring to light several 
metrics which correspond to classically integrable dynamical systems. They 
include, as particular cases, the Eguchi-Hanson and the Taub-NUT metrics.
\end{abstract}
\end{titlepage}

\section{Introduction}
The discovery of the generalized Runge-Lenz vector for the Taub-NUT  
metric \cite{gm} has been playing an essential role in the analysis of its 
classical and quantum dynamics. As shown in \cite{fh} this triplet 
of conserved quantities gives quite 
elegantly the quantum bound states as well as the scattering states.  
The Killing-St\" ackel tensors, which are the roots of the 
generalized Runge-Lenz vector, have been derived in \cite{gr} using 
purely geometric tools. As a result the classical integrability 
of the Taub-NUT metric was established. The other important metric of Eguchi-Hanson 
escaped to such an analysis (even though the results obtained in \cite{Mi} suggested 
strongly classical integrability), to say nothing of the full family of the 
multi-centre metrics. It is the aim of this article to fill this gap.

In section 2 we have gathered a summary of known properties of the 
multi-centre metrics, their geodesic flow and some basic concepts about 
Killing-St\" ackel and Killing-Yano tensors.

In section 3 we obtain the most general structure of the conserved quantity 
associated to a Killing-St\" ackel tensor: it is a bilinear form in the momenta. 
Taking this quadratic structure as a starting point, we obtain the 
system of equations which ensure 
that such kind of a quantity is preserved by the geodesic flow. This system is 
analyzed and simplified. Its most important consequence is that the 
existence of an extra conserved quantity is related to the existence of an 
extra spatial Killing (besides the tri-holomorphic one), which 
may be either holomorphic or tri-holomorphic.

In section 4 we first consider the case of an extra spatial Killing which 
is holomorphic. We find that the extra conserved quantity does exist for the 
following families, with isometry $U(1)\times U(1)$:
\brm
\item The most general two-centre metric, with the potential
\[V=v_0+\frac{m_1}{|\vec{r}+\vec{c}|}+\frac{m_2}{|\vec{r}-\vec{c}|},\]
which includes the double Taub-NUT metric for real $\,m_1=m_2$ and the 
Eguchi-Hanson metric when we have in addition $\,v_0=0.$ Our approach explains quite 
simply why there are three extra conserved quantities for Taub-NUT and only 
one for Eguchi-Hanson, and their very different nature.
\item A first dipolar breaking of Taub-NUT, with potential
\[V=v_0+\frac mr+{\cal F}\,\frac z{r^3}.\]
In the Taub-NUT limit ${\cal F}\to 0$ the extra conserved quantity becomes trivial.
\item A second dipolar breaking of Taub-NUT with potential
\[V=v_0+\frac mr+{\cal E}\,z.\]
In the Taub-NUT limit ${\cal E}\to 0$ there appears a triplet of extra 
conserved quantities: the generalized Runge-Lenz vector of \cite{gm}.

The classical integrability of these three dynamical systems follows from 
our analysis.
\erm

In section 5 we consider the case of an extra spatial Killing which is 
tri-holomorphic. We find four different families of metrics, which 
share with the previous ones their classical integrability. Some conclusions 
are presented in section 6.

\section{The Multi-Centre metrics}
\subsection{Background material}
These euclidean metrics on $M_4$ have at least one 
Killing vector $\,\wdt{\cal K}=\pt_t\,$ and have the local form
\beq\label{mc1}
g=\frac 1V\,(dt+\T)^2+V\,\ga,\qq V=V(x),\qq\T=\T_i(x)\,dx^i,
\eeq
where the $x^i$ are the coordinates on $\ga.$ They are solutions of the empty 
space Einstein equations provided that :
\brm
\item The three dimensional metric $\ga$ is flat. Using cartesian coordinates $x^i$ 
we can write
\beq\label{mc2}
\ga=d\vec{x}\cdot d\vec{x}.\eeq
\item Some monopole equation
\beq\label{mc3}
dV=\eta\stg_{\ga}\,d\T \qq\qq (\eta=\pm 1).\eeq
\erm
Notice that the integrability condition for the monopole equation is $\,\De V=0,$ 
hence these metrics display an exact linearization of the empty space Einstein 
equations. 
They have been derived in many ways \cite{ksd},\cite{gh},\cite{Hi1},\cite{Hi2}. 
In this last reference the geometric meaning of the cartesian coordinates $x_i$ was 
obtained: they are nothing but the momentum maps of the complex 
structures under the circle action of $\pt_t.$

Let us summarize some background knowledge on the multi-centre metrics for further 
use. Taking for vierbein
\[E_a\ :\qq\quad E_0=\frac 1{\sqrt{V}}(dt+\T),\qq\qq E_i=\sqrt{V}\,dx_i\]
and defining as usual the spin connection $\,\Om_{ab}\,$ and the 
curvature $\,R_{ab}\,$ by
\[d E_a+\Om_{ab}\wedge E_b=0,\qq\quad R_{ab}=d\Om_{ab}+\Om_{as}\wedge \Om_{sb},\]
one can check that these metrics have a spin connection with $\,\eta-$self-duality: 
\[\Om^{(-\eta)}_i\equiv \Om_{0i}-\frac{\eta}{2}\,\eps_{ijk}\,\Om_{jk}=0,\qq
\Longrightarrow\qq R^{(-\eta)}_i=0,\] 
which implies the $\,\eta-$self-duality of their curvature. It follows that they 
are hyperk\" ahler and hence Ricci-flat.

The complex structures are given by the triplet of 2-forms
\beq\label{f7}
J_i=E_0\wedge E_i-\frac{\eta}{2}\,\eps_{ijk}\,E_j\wedge E_k=(dt+\T)\wedge dx_i-
\frac{\eta}{2}\, V\,\eps_{ijk}\,dx_j\wedge dx_k, \eeq
which are closed, in view of the hyperk\" ahler property of these metrics. 

Let us note that the self-duality of the complex structures and of the spin 
connection are opposite and that the Killing vector $\,\pt_t\,$ is tri-holomorphic. 

It is useful to define the Killing 1-form, dual of the vector $\wdt{\cal K}=\pt_t,$ 
which reads 
\beq\label{f8}
{\cal K}=\dst\frac{dt+\T}{V},\eeq
and plays some role in some in characterizing the multi-centre metrics. 

\newpage
\noindent Among these characterizations let us mention:
\brm
\item For the multi-centre metrics the differential $d{\cal K}$ has a self-duality 
opposite to that of the connection. A proof using spinors may be found in \cite{tw} 
and without spinors in \cite{gd}.
\item The multi-centre metrics possess at least one tri-holomorphic Killing. For 
a proof see \cite{gr}.
\erm

\nc{\dt}{\dot{t}}   \nc{\dx}{\dot{x}}    \nc{\dy}{\dot{y}}    \nc{\dep}{\dot{p}}

\subsection{Geodesic flow}
The geodesic flow is the Hamiltonian flow of the metric considered as a function on 
the cotangent bundle of $M_4.$
Using the coordinates $(t,x_i)$ we will write a cotangent vector as
\[\Pi_i\,dx_i+\Pi_0\,dt.\]
The symplectic form is then
\beq\label{gf1}
\om=dx_i\wedge d\Pi_i+dt\wedge d\Pi_0,\eeq 
and we take for hamiltonian
\beq\label{gf2}
H=\frac 12\,g^{\mu\nu}\,\Pi_{\mu}\,\Pi_{\nu}=
=\frac 12\left(\frac 1V\,(\Pi_i-\Pi_0\,\T_i)^2+V\,\Pi_0^2\right).\eeq
For geodesics affinely parametrized by $\,\la\,$ the equations for 
the flow allow on the one hand to express the velocities
\beq\label{gff3}\barr{l}
\dst\dot{t}\equiv\frac{dt}{d\la}=\frac{\pt H}{\pt \Pi_0}=
\left(V+\frac{\T^2}{V}\right)\Pi_0-\frac{\T_i\Pi_i}{V},\\[4mm]
\dst\dot{x}_i\equiv\frac{dx_i}{d\la}=\frac{\pt H}{\pt p_i}=\frac 1V\,p_i,
\qq\quad p_i=\Pi_i-\Pi_0\,\T_i,
\earr\eeq
and on the other hand to get the dynamical evolution equations
\beq\label{gf3}\barr{l}\dst 
\dot{\Pi}_0=-\frac{\pt H}{\pt t}=0,
\qq\qq\qq \Pi_0=\frac{(\dt+\T_i\,\dx_i)}{V},\hfill (a)\\[4mm]\dst 
\dot{\Pi}_i=-\frac{\pt H}{\pt x_i}\qq\quad\Longrightarrow\qq \quad
\dot{p}_i=\left(\frac HV-q^2\right)\,\pt_i V+\frac qV\,(\pt_i\T_s-\pt_s\T_i)\,p_s.
\qq\hfill(b)
\earr\eeq
Relation (\ref{gf3}a) expresses the conservation of the charge 
$\dst\,q=\Pi_0,$ a consequence of the $U(1)$ isometry of the metric.


The conservation of the energy 
\beq\label{gf5}
H=\frac 12\left(\frac{p_i^2}{V}+q^2\,V\right)=\frac V2(\dx_i^2+q^2)=
\frac 12\,g_{\mu\nu}\,\dx^{\mu}\,\dx^{\nu}\eeq
is obvious since it expresses the constancy of the length of the tangent vector 
$\,\dx^{\mu}\,$ along a geodesic.

For the multi-centre metrics, use of relation (\ref{mc3}) brings the 
equations of motion to the nice form  
\beq\label{gf6}
\dot{\vec{p}}=\left(\frac HV-q^2\right)\,\vec{\nabla}\,V+
\eta\frac qV\ \vec{p}\wedge\vec{\nabla} V.
\eeq

\subsection{Killing-St\" ackel versus Killing-Yano tensors}
A Killing-St\" ackel (KS) tensor is a symmetric tensor $\,S_{\mu\nu}\,$ which satisfies
\beq\label{ks1}
\nabla_{(\mu}\,S_{\nu\rho)}=0.\eeq
Let us observe that if $\,K\,$ and $\,L\,$ are two (possibly different) Killing 
vectors their symmetrized tensor product $\,K_{(\mu}\,L_{\nu)}\,$ 
is a KS tensor. So we will define {\em irreducible} KS tensors as the ones which 
cannot be written as linear combinations, with constant coefficients, of 
symmetrized tensor products of Killing vectors. 

For a given KS tensor $\,S_{\mu\nu}\,$ the quantity
\beq\label{ks2}
{\cal S}=S_{\mu\nu}\,\dx^{\mu}\,\dx^{\nu}\eeq
is preserved by the geodesic flow. 

A Killing-Yano (KY) tensor is an antisymmetric tensor $\,Y_{\mu\nu}\,$ which satisfies
\beq\label{ks3}
\nabla_{(\mu}\,Y_{\nu)\rho}=0.\eeq
For instance a complex structure is an obvious KY tensor.  
One can build KS tensors from KY using:
\begin{nth}
If $\,Y\,$ and $\,Z\,$ are Killing-Yano tensors, then 
the tensor $\,S_{\mu\nu}=Y_{\mu}^{~\si}\,Z_{\si\nu}+Z_{\mu}^{~\si}\,Y_{\si\nu}\,$ 
is Killing-St\" ackel.\end{nth}

In \cite{gr} the triplet of KS tensors for the Taub-NUT metric was 
constructed, using proposition 1, from the known triplet of complex 
structures and a newly discovered KY tensor. 

\section{Quadratic conserved quantities}
Instead of focusing ourselves on the KS tensor $\,S_{\mu\nu},$ , whose usefulness 
is just to produce the conserved quantity ${\cal S},$ let us rather examine more 
closely the structure of the  conserved quantity itself. To this end we 
expand relation (\ref{ks2}) and get rid of the velocities using 
$\,\dt=qV-\T_i\,\dx_i\,$ and $\,\dst \dx_i=\frac{p_i}{V}.$ This gives, for the  
conserved quantity we are looking for, the structure
\beq\label{ks4}
{\cal S}=A_{ij}(x)\,p_i\,p_j+2q\,B_i(x)\,p_i+C(x).\eeq

Our starting point will be to look for a conserved quantity having this 
particular property of being quadratic with respect to the momenta, forgetting 
that it was generated by a Killing-St\" ackel tensor. Using the equations 
of the geodesic flow (\ref{gff3}),(\ref{gf3}) it is straightforward to 
impose that ${\cal S}$ be conserved. One gets:

\begin{nth}
The quantity $\,{\cal S}\,$ is conserved iff the following equations are satisfied
 \footnote{We want to obtain quantities conserved for any values of $q$ and $H.$ 
So we have $\,H-q^2\,V\neq 0.$}
\beq\label{ks}\barr{l}\dst 
\dl_B\,V=0 \hfill (a)\\[4mm]
\pt_{(k}\,A_{ij)}=0 \hfill (b)\\[4mm]
\pt_{(i}\,B_{j)}-\eta A_{s(i}\,\eps_{j)su}\,\pt_uV=0 \hfill (c)\\[4mm]
\pt_i\,C+2(H-q^2\,V)\,A_{is}\,\pt_s\,V-2\eta\,q^2\,\eps_{ist}\,B_s\,\pt_tV=0 
\qq\hfill (d)\earr\eeq
\end{nth}

Relation (\ref{ks}a) shows that, to have some extra conservation law, we need an 
extra symmetry for the potential function V. This means that $\,B_i\,$ must 
be conformal to some spatial Killing vector $\,K_i,$ which is a symmetry 
of the potential $\,V.$ Of course $\,K_i\,$ lifts up to an isometry of 
the 4 dimensional metric. So we have obtained:

\begin{nth}
The number of extra conserved quantities  of a multi-centre metric is at 
most equal to the number of extra {\rm spatial} Killing vectors it does 
possess (besides the tri-holomorphic Killing $\,\wdt{\cal K}=\pt_t$).
\end{nth}

\noindent Using this result we can discuss the triaxial generalization of 
the Eguchi-Hanson metric, with a tri-holomorphic $su(2),$ discovered in \cite{bgpp}. 
Its potential and cartesian coordinates were given in \cite{gorv} in terms of 
the usual spherical coordinates. From these results it follows that this metric 
has no spatial Killing vector hence it will have no (irreducible) KS tensor.

For further analyses it is useful to define
\beq\label{ks5}
B_i=-\eta\,F\,K_i.\eeq  
The conserved quantity (\ref{ks4}) becomes
\beq\label{cq}
{\cal S}=A_{ij}(x)\,p_i\,p_j-2\eta\,q\,F\,K_i\,p_i+C(x),\eeq
and equation (\ref{ks}c) transforms into
\beq\label{ks6}
K_{(i}\,\pt_{j)}F+A_{s(i}\,\eps_{j)su}\pt_u\,V=0.\eeq
Taking its trace we see that $\,\dst\dl_K\,F=0,$ showing that $\,V\,$ and 
$\,F\,$ must have the same Killing.

\subsection{Transformations of the system}
Contracting (\ref{ks6}) with $\pt_jV$ gives 
\begin{nlem}\label{l1}
The equation (\ref{ks6}) has for consequence:
\beq\label{ksup}
\quad (dV\cdot dF)K+\star(A[dV]\wedge dV)=0.\eeq
\end{nlem}
We can proceed to:
\begin{nth}
The relation (\ref{ks6}) is equivalent (except possibly at the points where 
the norm of the Killing $K$ vanishes) to the relations:
\beq\label{ks7}\left\{\barr{l}
A[K]=a(x)\,K, \hfill a)\\[4mm]
|K|^2\,dF-A[\star(K\wedge dV)]+\star(A[K]\wedge dV)=0\qq\qq\hfill b)
\earr\right.\eeq
\end{nth}

\noindent{\bf Proof :} Contracting relation (\ref{ks6}) with $K_j$ gives 
relation b), while contracting with $K_iK_j$ we have
\beq\label{kss1}
\eps_{stu}K_s A[K]_t \pt_u V=0\qq\Longrightarrow\qq A[K]_i=a(x)K_i+b(x)\pt_i V,\eeq
which is not relation a). To complete the argument we first contract 
relation (\ref{ks6}) with $\eps_{iab}K_a$; after some algebra we get
\beq\label{kss2}
K_j\eps_{iab}\pt_iF K_a+2A[K]_j\pt_b V+A[dV]_b K_j
-K_s A[dV]_s\,\de_{jb}-A_{ss}K_j\pt_bV=0,\eeq
which, upon contraction with $A[K]_b,$ gives eventually
\beq\label{kss3}
(A[K]_s\pt_sV)\,A[K]_i=\{-\eps_{stu}K_s A[K]_t \pt_u F+A_{ss}\,A[K]_t\pt_t V
-A[K]_s A[dV]_s\}\,K_i.\eeq
Let us now suppose that $A[K]_s\pt_s V\neq 0.$ The previous relation shows that in 
(\ref{kss1}) we must have $b(x)=0,$ hence $A[K]_s\pt_s V=0$ which 
is a contradiction. 

\noindent Let us prove that the converse 
is true. From $(\ref{ks7}b)$ we get
\beq\label{kst1}
|K|^2\,K_{(i}\,\pt_{j)}F+(K_{(j}A_{i)s}\,K_t\eps_{tsu}+
A[K]_s\,K_{(j}\eps_{i)su})\pt_u V=0.\eeq
Use of the identity
\beq\label{kst2}
A_{is}K_t\,K_j\eps_{tsu}\pt_u V=(|K|^2\,A_{is}\eps_{jsu}-A[K]_i K_t \eps_{jtu})
\pt_u V\eeq
and of relation $(\ref{ks7}a)$ leaves us with $(\ref{ks6}),$ up to division by   
$\,|K|^2.$ Notice that $\,|K|^2\,$ vanishes at the fixed points under the 
Killing action, i. e. in subsets of zero measure in ${\mb R}^3.$ \black

\noindent We can give, using (\ref{ks7}a) and the identity
\beq\label{id1}
-A[\star(K\wedge dV)]=\star(A[K]\wedge dV)-A_{ss}\,\star(K\wedge dV)
+\star(K\wedge A[dV]),\eeq
a simpler form to the relation (\ref{ks7}b):

\begin{nlem}
The relation (\ref{ks7}b) is equivalent to
\beq\label{ksup3}
|K|^2\,dF+(2a-{\rm Tr}\,A)\,\star(K\wedge dV)+\star(K\wedge A[dV])=0.\eeq
\end{nlem}
For further use let us prove:

\begin{nlem}\label{l2} 
To the spatial Killing $\,K,$ leaving the potential $V$ invariant, there 
corresponds a quantity $\,Q\,$ invariant under the geodesic flow given by
\beq\label{sup1}
Q=K_i\,p_i+\eta qG,\qq\mbox{with}\qq i(K)F=-\eta\,dG.\eeq
\end{nlem}

{\bf Proof :} We start from $\,\dst\dl_K\,V=0.$ Since $K$ is a Killing we have 
$\,\dst\dl_K(\star\,dV)=\star\,d(\dl_K\,V)=0,$ and (\ref{mc3}) implies that  
$\,\dst\dl_K\,d\T=0.$ The closedness of $\,d\T\,$ implies $\,d(i(K)d\T)=0,$ and 
since our analysis is purely local in $\,{\mb R}^3,$ we can define 
\beq\label{ksup2}
\,\eta\,dG=-i(K)\,d\T,\qq \Longrightarrow \qq \star(K\wedge dV)=dG.\eeq
Then we multiply (\ref{gf3}b) by $\,p_i$ and get successively
\[K_i\dep_i=\dot{(K_ip_i)}-\dot{K}_i\,p_i=\dot{(K_ip_i)}=
\frac qV K_i\,(\pt_i\T_s-\pt_s\T_i)p_s
=-\eta q\,\dx_s\pt_s G=-\eta q\,\dot{G},\]
which concludes the proof. \black

\noindent Let us point out that if we use the coordinate $\phi$ adapted to 
the Killing $\tilde{K}=\pt_{\phi},$ we can write the connection $\,\T=\eta G\,d\phi,$
 where $\,G\,$ does not depend on $\,\phi.$

\subsection{Integrability equations}
We will derive now the integrability conditions for the equations (\ref{ks}c) and 
(\ref{ks}d). The first one was written using forms in (\ref{ksup3}) while the 
second one is
\beq\label{extra1}
dC+2(H-q^2 V)A[dV]+2q^2 F\,\star(K\wedge dV)=0.\eeq
It can now be proved :

\begin{nth} The integrability condition for (\ref{extra1}) is 
\beq\label{ks8}
d\,A[dV]=0\qq\qq\Longrightarrow\qq A[dV]=dU\qq\mbox{and}\qq \dl_K U=0.\eeq 
\end{nth}

{\bf Proof :} The integrability condition is obtained by 
differentiating (\ref{extra1}). We get
\beq\label{ks9}
2(H-q^2V)\,d\,A[dV]+2q^2\,A[dV]\wedge dV+2q^2\,dF\wedge\star(K\wedge dV)
+ 2q^2F\,d\star(K\wedge dV)=0.\eeq
The last term in this equation vanishes in view of (\ref{ksup2}). Furthermore 
we have the identity specific to three dimensional spaces
\[dF\wedge\star(K\wedge dV)=-(K\cdot dF)\,\star dV+(dV\cdot dF)\,\star K
=(dV\cdot dF)\,\star K\]
because $K$ is a symmetry of $F.$ Relation (\ref{ks9}) simplifies to
\[2(H-q^2 V)\,d\,A[dV]+2q^2\,\star[(dV\cdot dF)\,K+\star(A[dV]\wedge dV)]=0,\]
and lemma \ref{l1} implies the closedness of $A[dV].$ 
Since our analysis is purely local, the existence of $U$ is a consequence of  
Poincar\'e's lemma. 

The relations 
\[\dl_K U=i(K)\,dU=i(K)\,A[dV]=(A[K]\cdot dV)=a(K\cdot dV)=a\,\dl_K V=0\]
show the invariance of $U$ under the Killing $K.$ \black

Let us now turn to equation (\ref{ksup3}). We will prove:

\begin{nth}\label{prop1}
The integrability condition for (\ref{ksup3}) is
\beq\label{ks10}
(2a-{\rm Tr}\,A)dV+dU=|K|^2\,\star d\tau,\qq\qq \dl_K\,d\tau=0,\eeq
for some one form $\tau.$
\end{nth}

{\bf Proof :} Let us define the 1-form
\beq\label{pr1}
Y=(2a-{\rm Tr}\,A)dV+dU.\eeq
It allows to write (\ref{ksup3}) and its integrability condition as
\beq\label{pr2}
dF=-\star\left(\frac{K\wedge Y}{|K|^2}\right),\qq\qq 
\de\left(\frac{K\wedge Y}{|K|^2}\right)=0,\eeq
or switching to components
\beq\label{pr4}
K_i\,\de\left(\frac{Y}{|K|^2}\right)+\frac{Y_s\pt_s K_i-K_s\pt_sY_i}{|K|^2}=0.\eeq
Let us examine the last terms. Since $a$ and ${\rm Tr}\,A$ are invariant under the 
Killing $K,$ we obtain
\beq\label{pr5}
Y_s\pt_sK_i-K_s\pt_sY_i=-(2a-{\rm Tr}\,A)\pt_i(K_s\pt_s\,V)
-\pt_i(K_s\pt_s\,U)\eeq
and both terms vanish because $V$ and $U$ are invariant under $K.$ We are left with 
the vanishing of the divergence of $Y/|K|^2$ from which we conclude (local analysis!) 
that it must have the structure $\,\star d\tau$ for some 1-form $\tau.$ >From its 
definition it follows that $\,d\tau$ is invariant under $\,K.$ \black

Using this result we can simplify (\ref{ksup3}) to 
\beq\label{ks11}
dF+\star(K\wedge\star d\tau)=dF-i(K)d\tau=0.\eeq
Collecting all these results we have:
 
\begin{nth}
The quantity
\[{\cal S}=A_{ij}(x)\,p_i\,p_j-2\eta\,q\,F\,K_i\,p_i+C(x)\]
is preserved by the geodesic flow of the multi-centre metrics provided that 
the integrability constraints
\beq\label{ksf2}
\De\,V=0,\qq A[dV]=dU,\qq (2a-{\rm Tr}\,A)\,dV+dU=|K|^2\,\star d\tau\eeq
and the following relations hold:
\beq\label{ksf1}\barr{l}\dst 
\dl_K V=0,\\[4mm]
\pt_{(k}A_{ij)}=0,\qq\qq  A[K]=a\,K, \\[4mm]
dF=i(K)\,d\tau,\\[4mm] 
d(C+2HU)+2q^2(-V\,dU+F\,dG)=0,\qq\qq \star(K\wedge dV)=dG.
\earr\eeq
\end{nth}

\subsection{Classification of the spatial Killing vectors}
An important point, in view of classification, is 
whether the extra spatial Killing is tri-holomorphic 
or not. This can be checked thanks to:
\begin{nlem}\label{l3}
The spatial Killing vector $\,K_i\pt_i\,$ is tri-holomorphic iff 
\[\eps_{ist}\pt_{[s}K_{t]}=0.\]
Otherwise it is holomorphic.
\end{nlem}

{\bf Proof :} From \cite{bf} we know that, for an hyperk\" ahler geometry, a 
Killing may be either holomorphic or tri-holomorphic. As shown in \cite{gr} 
such a vector will be tri-holomorphic iff the differential of the dual 1-form 
$\,K=K_i\,dx_i\,$ has the self-duality opposite to that of the complex 
structures. A computation shows that this is equivalent to the vanishing of
\[dK^{(-\eta)}=-\frac{\eta}{2}\,\eps_{ijk}\,\pt_{[j}\,K_{k]}
\left(E_0\wedge E_i-\frac{\eta}{2}\,\eps_{ist}\,E_s\wedge E_t\right),\]
from which the lemma follows. \black

Since we are working in a flat three dimensional flat space, there are 
essentially two different cases to consider:
\brm
\item The Killing $K$ generates a spatial rotation, which we can take, without 
loss of generality, around the z axis. In this case we have
\[K_i\,p_i=L_z\]
and this Killing vector is holomorphic with respect to the complex structure 
$\,J_3,$ defined in section 2.
\item The Killing $K$ generates a spatial translation, which we can take, without 
loss of generality, along the z axis. In this case we have the 
\[K_i\,p_i=p_z\]
and this Killing vector is tri-holomorphic.
\erm

We will discuss successively these two possibilities.

\section{One extra holomorphic spatial Killing vector}
One can get the general solution of the first equation for $\,A_{ij}\,$ in 
(\ref{ksf1}b). It is most conveniently written in terms of 
$\,{\cal A}(p,p)\equiv A_{ij}\,p^i\,p^j.$ One has:
\beq\label{rk1}
{\cal A}(p,p)=\left\{\barr{l}
\alf\,L_x^2+\be\,L_y^2+\ga\,L_z^2+2\mu\,L_y L_z +2\nu\,L_z L_x+2\la\,L_x L_y\\[4mm]
+a_1\,p_x L_y+a_2\,p_x L_z+b_1\,p_y L_x+b_2\,p_y L_z+c_1\,p_z L_x+c_2\,p_z L_y\\[4mm]
+d_1\,p_x L_x+d_2\,p_y L_y+a_{ij}p_ip_j.\earr\right.\eeq
At this point we have 20 free parameters. They reduce, when one imposes the 
existence of the rotational Killing, to
\beq\label{rk2}
{\cal A}(p,p)=\alf(L_x^2+L_y^2)+\ga\,L_z^2
+b\,(\vec{p}\wedge \vec{L})_z+a_{33}\,p_z^2+a_{11}\vec{p}\,^2+\de\,p_z L_z.\eeq
We note that the parameter $\,\ga\,$ corresponds to a {\em reducible} piece which 
is just the square of $\,L_z.$ We will take $\,\ga=\alf\,$ for convenience. 

The parameter $\,a_{11}\,$ is easily seen, upon integration 
of the remaining equations in (\ref{ks}), to give rise, in the conserved 
quantity $\,{\cal S},$ to the full piece
\beq\label{rk3}
a_{11}(\vec{p}\,^2-2HV+q^2V^2)\eeq
which vanishes thanks to the energy conservation (\ref{gf5}). So we can take 
$\,a_{11}=0.$ 

The second relation in (\ref{ksf1}b) implies the vanishing of $\,\de.$ Hence, 
with slight changes in the notation, we end up with \beq\label{rk6}
{\cal A}(p,p)=a\,\vec{L}\,^2+c^2\,p_z^2+b\,(\vec{p}\wedge\vec{L})_z.\eeq
Let us note that the parameters $a$ and $b$ are real while the parameter $c$ 
may be either real or pure imaginary.

To take advantage of the rotational symmetry around the z axis we use first the 
coordinates $\,\rho=x^2+y^2\,$ and $z.$ From the system (\ref{ksf1}) one can 
check that the functions $F$ and $U$ are to be determined from
\beq\label{fc1a}\left\{\barr{l}
F_{,\rho}=(az+b/2)V_{,z}-a/2\,V_{,z}\\[4mm] 
F_{,z}=2(az^2+bz-c^2)V_{,\rho}-(az+b/2)V_{,z}\earr\right.\eeq
and 
\beq\label{fc1b}\left\{\barr{l}
U_{,\rho}=z(az+b)V_{,\rho}-\frac 12(az+b/2)V_{,z}\\[4mm]
U_{,z}=-2\rho(az+b/2)V_{,\rho}+(a\rho+c^2)V_{,z}\earr\right.\eeq
We will write the connection 
\beq\label{con1}
\T=\eta \,G\,d\phi,\qq x=\sqrt{\rho}\,\cos\phi,\qq y=\sqrt{\rho}\,\sin\phi.\eeq
Using lemma \ref{l2} we get the conserved quantity 
\beq\label{cq1}
\,J_z=L_z+\eta q\,G=x\,\Pi_y-y\,\Pi_x,\eeq 
which will be useful in the integrability proof.

\subsection{The two-centre metric}
This case corresponds to the choice $\,a=1\,$ and $\,c\neq 0.$ 
Since $a=1,$ we can get rid of the constant $b$ by a 
translation of the variable $z.$ So, without loss of generality, we can take 
$b=0$ and use the new variables 
$\,r_{\pm}=\sqrt{x^2+y^2+(z\pm c)^2}.$ We get the relations
\[\pt_{r_+}F=-c\,\pt_{r_+} V,\qq\quad \pt_{r_-}F=+c\,\pt_{r_-} V\]
which imply
\[V=f(r_+)+g(r_-),\qq\qq F=-c(f(r_+)-g(r_-)).\]
Imposing to the potential $V$ the Laplace equation we have
\beq\label{fc2}
V=v_0+\frac{m_1}{r_+}+\frac{m_2}{r_-},
\qq\qq F=-c\left(\frac{m_1}{r_+}-\frac{m_2}{r_-}\right)=-c\De,\eeq
i. e. we recover the most general 2-centre metric. Let us recall that the 
double Taub-NUT metric, given by real $\,m_1=m_2,$ is complete. If in addition 
we take the limit $\,v_0\to 0,$ we are led to the Eguchi-Hanson \cite{eh} metric.

One has then to check the integrability constraint (\ref{ks8}) and to determine 
the functions $U$ and $C$ \footnote{We discard constant 
terms in the function $C.$}
\beq\label{fc3}
U=-cz\De,\qq\qq
C=-2(H-q^2\,V)U-q^2\,r^2\,\De^2,\qq r^2=x^2+y^2+z^2.\eeq
Let us observe that the conserved quantity which we obtain may be real even if 
$c$ is pure imaginary. In this case $m_1=m$ may be complex, but if we take 
$m_2=m^{\star}$ the functions $V$ and $c\De$ are real, as well as ${\cal S}.$

The final form of the conserved quantity for the two-centre metric is therefore
\beq\label{fc4}\left\{\barr{l}
{\cal S}_I=\vec{L}\,^2+c^2\,p_z^2+2\eta qc\,\De\,L_z+2cz\,\De\,(H-q^2V)
-q^2 r^2\De^2 \\[4mm]
\dst V=v_0+\frac{m_1}{r_+}+\frac{m_2}{r_-}
\qq\qq\De=\frac{m_1}{r_+}-\frac{m_2}{r_-} \earr\right.\eeq

\noindent This conserved quantity is certainly different of the one 
exhibited in \cite{gr} since the latter does trivialize for the 
Eguchi-Hanson case while the former does not.

\noindent For completeness let us give the connection:
\beq\label{fc5}
\T=\eta\,G\,d\phi,\qq\qq G=m_1\,\frac{z+c}{r_+}+m_2\,\frac{z-c}{r_-}.
\eeq

>From the very definition of the coordinates $r_{\pm}$ it is clear that 
the previous analysis is only valid for $c\neq 0.$ The 
special case $c=0$ will be examined now.

\subsection{First dipolar breaking of Taub-NUT}
This case corresponds to the choice $\,a=1\,$ and $\,c=0.$ 
Since $a=1,$ we can again get rid of the parameter $b.$ Then 
relation (\ref{fc1a}) for $F$ implies 
\beq\label{sc1}
V=w_0(r)+w_1(r)\,z,\qq\qq F_{,r}=-rw_1(r).\eeq
Imposing the Laplace equation we obtain
\beq\label{sc2}
V=v_0+\frac mr+{\cal E}z+{\cal F}\frac z{r^3},\qq\qq 
F=-\frac{\cal E}{2}\,r^2+\frac{\cal F}{r}.\eeq
The integrability relations for $U$ require that ${\cal E}=0$ and we have
\beq\label{sc3}
U={\cal F}\frac zr,\qq\qq C=-2{\cal F}\,\frac zr\,(H-q^2V)
-2mq^2{\cal F}\,\frac z{r^2}-q^2{\cal F}^2\,\frac{(3z^2-r^2)}{r^4}.\eeq

The final form of the conserved quantity is therefore
\beq\label{sc4}\left\{\barr{l}\dst
{\cal S}_{II}=\vec{L}\,^2-2\eta q\,\frac{\cal F}{r}\,L_z
-2{\cal F}\,\frac zr\,(H-q^2v_0)+q^2{\cal F}^2\frac{(x^2+y^2)}{r^4}\\[4mm]
\dst V=v_0+\frac mr+{\cal F}\frac z{r^3}\earr\right.\eeq
The connection is:
\beq\label{sc5}
\T=\eta\,G\, d\phi,\qq\qq G=m\frac zr-{\cal F}\,\frac{x^2+y^2}{r^3}.
\eeq
Let us note that in the Taub-NUT limit (${\cal F}\to 0$) the conserved quantity 
${\cal S}_{II}$ does become trivial.

\subsection{Second dipolar breaking of Taub-NUT}
This case corresponds to the choice $\,a=0\,$ and $\,b=1.$ 
The relation (\ref{fc1a}) shows that by a translation of $z$ we can take, without 
loss of generality, $c=0.$ From the integrability of $F$ we deduce
\beq\label{tc1}
V=f(r)+g(z),\qq\qq F=\frac 12(f(r)-g(z)).\eeq
Imposing Laplace equation yields
\beq\label{tc2}
V=v_0+\frac mr+{\cal E}z,\qq\qq F=\frac 12\left(\frac mr-{\cal E}z\right)\eeq
Then the integrability conditions for $U$ are satisfied and we obtain
\beq\label{tc3}
U=\frac{mz}{2r}-\frac{\cal E}{4}\,(x^2+y^2),\qq C=-2U(H-q^2 v_0)
-2q^2\,m{\cal E}\,\frac{(x^2+y^2)}{r}.\eeq

The final form of the conserved quantity is therefore
\beq\label{tc4}\left\{\barr{l}\dst
{\cal S}_{III}=(\vec{p}\wedge\vec{L})_z
-\eta q\,\left(\frac mr-{\cal E}\,z\right)L_z
-2U\,(H-q^2 v_0)-2q^2\,m{\cal E}\,\frac{(x^2+y^2)}{r} \\[4mm]\dst 
V=v_0+\frac mr+{\cal E}\,z\qq\qq 
U=\frac{mz}{2r}-\frac{\cal E}{4}\,(x^2+y^2)\earr\right.\eeq
The gauge field $\,\T\,$ is given by
\beq\label{tc5}
\T=\eta\,G\, d\phi,\qq\qq G=m\frac zr+\frac{\cal E}{2}(x^2+y^2).\eeq

For ${\cal E}=0$ we are back to the Taub-NUT metric. In this case the spatial 
isometries are lifted up from $u(1)$ to $su(2).$ As a result we have now 
three possible Killings to start with
\beq\label{ko}
K_i^{(1)}p_i=L_x\qq K_i^{(2)}p_i=L_y\qq K_i^{(3)}p_i=L_z\eeq
and we expect that the conserved quantity found above should be part of a triplet. 
The two missing conserved quantities can be constructed following the same 
route which led to ${\cal S}_{III}$ using the new available spatial 
Killings given by (\ref{ko}). We recover \beq\label{rg1}
\vec{\cal S}=\vec{p}\wedge\vec{L}-\eta q\,\frac mr\,\vec{L}
+m(q^2v_0-H)\frac{\vec{r}}{r},\qq{\cal S}_{III}({\cal E}=0)\equiv{\cal S}_z.\eeq
Lemma \ref{l2} lifts up $\,J_z,$ given by (\ref{cq1}), to a triplet 
of conserved quantities
\beq\label{rg2}
\vec{J}=\vec{L}+\eta q\,\frac mr\,\vec{r},\eeq
which allows to write
\beq\label{rg3}
\vec{\cal S}=\vec{p}\wedge\vec{J}+m(q^2v_0-H)\frac{\vec{r}}{r},\eeq
on which we recognize the generalized Runge-Lenz vector discovered by 
Gibbons and Manton \cite{gm}.

We have therefore obtained, for the three hamiltonians 
$H_I,\,H_{II}({\cal F}\neq 0)\,$ and $\,H_{III},$ corresponding respectively 
to the extra conserved quantities $\,{\cal S}_I,\,{\cal S}_{II}$ and 
$\,{\cal S}_{III},$  a set of four conserved quantities:
\[H,\quad q=\Pi_0,\quad J_z,\quad{\cal S},\]
which can be checked to be in involution with respect to the Poisson bracket.  
\newpage\noindent Hence we conclude to:

\begin{nth}
The three hamiltonians $H_I,\,H_{II}({\cal F}\neq 0)\,$ and $\,H_{III},$  defined 
above are integrable in Liouville sense.
\end{nth}
This includes, as a special case, the classical integrability of the 
Eguchi-Hanson metric.

\section{One extra tri-holomorphic spatial Killing vector}
This time we have for Killing $K_ip_i=p_z.$ Imposing this translational invariance 
and the constraint $A[K]\propto K$ restricts  
${\cal A}(p,p)$ to have the form
\beq\label{tk1}  
{\cal A}(p,p)=a\,L_z^2-2b\,p_xL_z+2c\,p_yL_z+\sum_{i,j=1}^2\,a_{ij}\,p_ip_j.\eeq
We have omitted a term proportional to $p_z^2$ since it is reducible.

The functions $\,F\,$ and $\,U,$ which depend only on the coordinates $x$ and $y,$ 
using the system (\ref{ksf1}), are seen to be determined by
\beq\label{th1}
\left\{\barr{l}
F_{,x}=A_{12}\,V_{,x}-A_{11}\,V_{,y}\\[4mm]
F_{,y}=A_{22}\,V_{,x}-A_{12}\,V_{,y}\earr\right.\qq
\left\{\barr{l}
U_{,x}=A_{11}\,V_{,x}+A_{12}\,V_{,y}\\[4mm]
U_{,y}=A_{12}\,V_{,x}+A_{22}\,V_{,y}\earr\right.\eeq
with
\beq\label{th2}
A_{11}=ay^2+2by+a_{11},\qq A_{22}=ax^2+2cx+a_{22},\qq A_{12}=-axy-bx-cy+a_{12}.\eeq
The connection and the conserved quantity related to $p_z$ will be
\beq\label{sup5}
\T=\eta\,G\,dz,\qq\qq\Pi_z=p_z+\eta\,q\,G.\eeq

\noindent In order to organize the subsequent discussion, let us observe:
\brm
\item For $\,a\neq 0,$ we may take $\,a=1.$ The spatial translations allow to 
take $\,b=c=0,$  and a rotation  $a_{12}=0$ as well. Hence we are left with
\[{\cal A}(p,p)=L_z^2+(a_{11}-a_{22})\,p_x^2+a_{22}(p_x^2+p_y^2).\]
Adding the reducible term $\,a_{22}\,p_z^2$ we recover the piece 
$a_{22}\,\vec{p}\,^2\,$ which can be discarded, as already explained in section 4. 
So we will take for our first case
\beq\label{thk1}
{\cal A}_1(p,p)=L_z^2-c^2\,p_x^2,\qq c\in{\mb R}\cup i{\mb R}, \qq c\neq 0.\eeq
\item Our second case, which is the singular limit $\,c\to 0\,$ of the first case, 
corresponds to
\beq\label{thk2}
{\cal A}_2(p,p)=L_z^2.\eeq
\item For $\,a=0,$ a first translation allows to take $\,a_{12}=0,$ while the 
second one allows the choice $\,a_{11}=a_{22}\,$ and the corresponding term 
$\,a_{11}(p_x^2+p_y^2)\,$ is disposed of as in the first case. Eventually a rotation 
will bring $b$ to zero and $c=1.$ Our third case will be
\beq\label{thk3}
{\cal A}_3(p,p)=p_y\,L_z.\eeq
\item For $a=b=c=0,$ a rotation brings $\,a_{12}\,$ to zero. Discarding 
$\,p_x^2+p_y^2,$ we are left with our fourth case
\beq\label{thk4}
{\cal A}_4(p,p)=p_x^2-p_y^2.\eeq
\erm
We will state the results obtained for these four cases without  
going through the detailed computations, which are greatly simplified using 
the complex coordinates $\,\z=x+iy\,$ and $\,\ol{\z}=x-iy.$

\subsection{First case}
Writing the corresponding metric 
\beq\label{Fc3}
\frac 1V(dt+\tau\,dz)^2+V(dz^2+d\ol{\z}\,d\z),\qq 
\eeq
and the conserved quantity as
\beq\label{Fc1}
{\cal S}_1=L_z^2-c^2\,p_x^2-2\eta q F\,\Pi_z+2(q^2v_0-H)\,U+q^2D,\qq
\qq c\neq 0,\eeq
we have
\beq\label{Fc2}\barr{l}\dst
\bullet\quad V=v_0+m\,\frac{\z}{\sqrt{\z^2+c^2}}+
\ol{m}\,\frac{\ol{\z}}{ \sqrt{\ol{\z}^2+c^2}},\qq v_0\in{\mb R},
\quad m\in{\mb C}\\[4mm]\dst
\bullet\quad V-v_0+i\,G=2m\frac{\z}{\sqrt{\z^2+c^2}},\qq 
U+iF=-mc^2\,\frac{\z+\ol{\z}}{\sqrt{\z^2+c^2}},\\[4mm]\dst
\bullet\quad D=-2c^2\,|m|^2\,\frac{(\z^2+\ol{\z}^2+|\z|^2+c^2)}{|\z^2+c^2|}.
\earr\eeq
If one is willing to use spheroidal coordinates $\,\xi\,$ and $\eta\,$ defined 
(for $c^2>0$) by
\[x=\frac 1c\,\sqrt{(\xi^2-c^2)(c^2-\eta^2)}\qq\qq y=\frac 1c\,\xi\eta,\]
it is possible to write the results in terms of real quantities. This 
has the drawback that the cases $\,c^2>0\,$ and $\,c^2<0\,$ need a separate analysis.

\subsection{Second case}
Writing the conserved quantity as
\beq\label{Sc1}
{\cal S}_2=L_z^2-2\eta qF\,\Pi_z+2(q^2v_0-H)\,U,\eeq
we have:
\beq\label{Sc2}\barr{l}\dst
\bullet\quad V=v_0+\frac{m}{\z}+\frac{\ol{m}}{\ol{\z}},
\qq v_0\in{\mb R},\quad m\in{\mb C},\\[4mm]\dst
\bullet\quad V-v_0+i\,G=2\,\frac m{\z},\qq 
U+iF=2m\,\frac{\ol{\z}}{\z}.
\earr\eeq

\subsection{Third case}
Writing the conserved quantity as
\beq\label{Tc1}
{\cal S}_3=p_y\,L_z-2\eta qF\,\Pi_z+2(q^2v_0-H)\,U+q^2D,\eeq
we have:
\beq\label{Tc2}\barr{l}\dst
\bullet\qq V=v_0+\frac{m}{\sqrt{\z}}+\frac{\ol{m}}{\sqrt{\ol{\z}}},
\qq v_0\in{\mb R},\quad m\in{\mb C},\\[4mm]\dst
\bullet\quad V-v_0+i\,G=2\,\frac m{\sqrt{\z}},\qq 
U+iF=\frac m2\,\frac{\ol{\z}-\z}{\sqrt{\z}}\\[4mm]\dst
\bullet\qq D=|m|^2\left( \sqrt{\frac{\ol{\z}}{\z}}+\sqrt{\frac{\z}{\ol{\z}}}\right)
\earr\eeq

\subsection{Fourth case}
Writing the conserved quantity as
\beq\label{Ffc1}
{\cal S}_4=p_x^2-p_y^2-2\eta qF\,\Pi_z+2(q^2v_0-H)\,U+q^2 D\eeq
we have
\beq\label{Ffc2}\barr{l}
\bullet\qq V=v_0+m\,\z+\ol{m}\,\ol{\z},\qq v_0\in{\mb R},\quad m\in{\mb C},\\[4mm]
\bullet\qq V-v_0+i\,G=2m\,\z,\qq U+iF=2m\,\ol{\z},\\[4mm]
\bullet\qq D=2|m|^2(\z^2+\ol{\z}^2).\earr\eeq

As was the case when the extra spatial Killing was holomorphic, we have obtained 
for the four hamiltonians considered in this section, a set of four conserved 
quantities
\[H,\quad q=\Pi_0,\quad \Pi_z,\quad {\cal S},\]
which are in involution with respect to the Poisson bracket, hence we conclude to:

\begin{nth}
The four hamiltonians determined in this section are integrable in Liouville sense.
\end{nth}
Let us conclude with two remarks:
\brm
\item One should notice that, among the four potentials considered in this section, 
only the second one and the fourth one are {\em uniform} functions in the 
three dimensional flat space.
\item The fourth case analyzed in this section is also interesting because it 
may exhibit {\em super-integrability}, i. e. more than four conserved quantities. 
According to the choice of the parameter $m$ we have one more spatial Killing 
and one more conserved quantity
\beq\label{FFc4}
m=\ol{m}\quad\Longrightarrow\quad \{\pt _y\ ,\ \Pi_y\}
\qq ; \qq m=-\ol{m}\quad\Longrightarrow\quad \{\pt_x\ ,\ \Pi_x\}.\eeq
The algebra generated by these five conserved quantities, with respect to the 
Poisson bracket, remains fully abelian. This phenomenon of super-integrability, 
which is of quite common experience in classical mechanics, seems quite rare 
within the multi-centre, since it is exhibited only by one family out of seven.
\erm

\section{Conclusion}
We have settled the problem of finding all the multi-centre metrics which do 
exhibit some extra quantity, quadratic with respect to the momenta, and preserved 
by the geodesic flow. The concept of Killing-St\" ackel can be generalized to 
of type $\,(n,0),$ with $\,n\geq 3.$ Such a tensor has to be fully symmetric 
and such that
\[\nabla_{(\la}S_{\mu_1\cdots\mu_n)}=0.\]
It follows that the geodesic flow preserves the quantity
\[S_{\mu_1\cdots\mu_n}\,\dx^{\mu_1}\cdots\dx^{\mu_n}.\]
The corresponding invariants will be cubic, quartic, etc... with respect to 
the momenta. Little is known about the existence of such objects for the multi-centre 
metrics.

Let us put emphasis also on the purely local nature of our analysis: it makes no 
difference between complete and non-complete metrics. For instance in section 4 
we have seen that the most general two-centre metric is integrable, however 
it is complete only for real $\,m_1=m_2,$ i. e. for the double Taub-NUT metric.



\end{document}